\documentclass[a4paper,11pt]{article}

\usepackage{amssymb,bbold}
\usepackage{amsmath}
\usepackage{authblk}                    
\usepackage[usenames,dvipsnames]{color}
\usepackage{tensor}                     
\usepackage{bbm}                        
\usepackage{graphicx}
\usepackage{cancel}
\usepackage{mathrsfs}
\usepackage{bm}
\usepackage{cancel}
\usepackage{setspace}                  
\usepackage[left=1in,right=1in,top=1in,bottom=1in]{geometry}                  

\usepackage{hyperref}
\hypersetup{
    colorlinks=true,         
    linkcolor=blue,          
    citecolor=red,        
    urlcolor=Violet             
}

\title{\bf Right about time?}

\author[1,2]{Sean Gryb\thanks{s.gryb@hef.ru.nl}}

\author[3]{Flavio Mercati\thanks{fmercati@perimeterinstitute.ca}}

\affil[1]{{\it Institute for Theoretical Physics}, Utrecht University\authorcr Leuvenlaan 4, 3584 CE Utrecht, The Netherlands}

\affil[2]{{\it Institute for Mathematics, Astrophysics and Particle Physics}, Radboud University\authorcr Huygens Building, Heyendaalseweg 135, 6525 AJ Nijmegen, The Netherlands}

\affil[3]{{\it Perimeter Institute for Theoretical Physics}\authorcr 31 Caroline Street North, Waterloo, ON N2L 2Y5, Canada.}

\date{\vspace{-36pt}}

\let\oldmarginpar\marginpar
\renewcommand\marginpar[1]{\oldmarginpar{\color{red}\raggedright\scriptsize #1}}
\newcommand{\eq}[1]{(\ref{eq:#1})}

\begin{document}

\maketitle
\begin{abstract}
    Have our fundamental theories got time right? Does size really matter? Or is physics all in the eyes of the beholder? In this essay, we question the origin of time and scale by reevaluating the nature of measurement. We then argue for a radical scenario, supported by a suggestive calculation, where the flow of time is inseparable from the measurement process. Our scenario breaks the bond of time and space and builds a new one: the marriage of time and scale.
\end{abstract}

\section{Introduction}

Near the end of the 19$^\text{th}$ century, physics appeared to be slowing down. The mechanics of Newton and others rested on solid ground, statistical mechanics explained the link between the microscopic and the macroscopic, Maxwell's equations unified electricity, magnetism, and light, and the steam engine had transformed society. But the blade of progress is double edged and, as more problems were sliced through, fewer legitimate fundamental issues remained. Physics, it seemed, was nearing an end.

Or was it? Among the few remaining unsolved issues were two experimental anomalies. As Lord Kelvin allegedly announced:``The beauty and clearness of the dynamical theory [...] is at present obscured by two clouds.''\cite{Kelvin:dark_clouds} One of these clouds was the ultra--violet catastrophe: an embarrassing prediction that hot objects like the sun should emit \emph{infinite} energy. The other anomaly was an experiment by Michelson and Morley that measured the speed of light to be independent of how an observer was moving. Given the tremendous success of physics at that time, it would have been a safe bet that, soon, even these clouds would pass. 

Never bet on a sure thing. The ultra--violet catastrophe led to the development of quantum mechanics and the Michelson--Morley experiment led to the development of relativity. These discoveries completely overturned our understanding of space, time, measurement, and the perception of reality. Physics was not over, it was just getting started.

Fast--forward a hundred years or so. Quantum mechanics and relativity rest on solid ground. The microchip and GPS have transformed society. These frameworks have led to an understanding that spans from the microscopic constituents of the nucleus to the large scale structure of the Universe. The corresponding models have become so widely accepted and successful that they have been dubbed \emph{standard models} of particle physics and cosmology. Resultantly, the number of truly interesting questions appears to be slowly disappearing. In well over 30 years, there have been no  experimental results in particle physics that can't be explained within the basic framework laid out by the standard model. With the ever increasing cost of particle physics experiments, it seems that the data is drying up. But without input from experiment, how can physics proceed? It would appear that physics is, again, in danger of slowing down.

Or is it? Although the \emph{number} of interesting fundamental questions appears to be decreasing, the \emph{importance} of the remaining questions is growing. Consider two of the more disturbing experimental anomalies. The first is the \emph{naturalness problem}, i.e., the presence of unnaturally large and small numbers in Nature. The most embarrassing of these numbers -- and arguably the worst prediction of science -- is the accelerated expansion of the Universe, which is some 120 orders of magnitude smaller than its natural value. The second is the \emph{dark matter problem} that just under 85--90 percent of the matter content of our Universe is of an exotic nature that we have not yet seen in the lab. It would seem that we actually understand very little of what is happening in our Universe!

The problem is not that we don't have enough data. The problem is that the data we do have does not seem to be amenable to explanation through incremental theoretical progress. The belief that physics is slowing down or, worse, that we are close to a final theory is just as as unimaginative now as it would have been before 1900. The lesson from that period is that the way forward is to question the fundamental assumptions of our physical theories in a radical way. This is easier said than done: one must not throw out the baby with the bath water. What is needed is a careful examination of our physical principles in the context of real experimental facts to explain \emph{more} data using \emph{less} assumptions.

The purpose of this work is to point out three specific assumptions made by our physical theories that might be wrong. We will not offer a definite solution to these problems but suggest a new scenario, supported by a suggestive calculation, that puts these assumptions into a new light and unifies them. The three assumptions we will question are
\begin{enumerate}
    \item Time and space are unified.
    \item Scale is physical.
    \item Physical laws are independent of the measurement process.
\end{enumerate}

We will argue that these three assumptions inadvertently violate the same principle: the requirement that the laws of physics depend only on what is knowable through direct measurement. They fall into a unique category of assumptions that are challenged when we ask how to adapt the scientific method, developed for understanding processes in the lab, to the cosmological setting. In other words, how can we do science on the Universe \emph{as a whole}?

We will not directly answer this question but, rather, suggest that this difficult issue may require a radical answer that questions the very origin of time. The flow of time, we will argue, may be fundamentally linked to the process of measurement. We will then support this argument with an intriguing calculation that recovers the black hole entropy law from a simple toy model. Before getting to this, let us explain the three questionable assumptions.

\section{Three questionable assumptions}

Many of our most basic physical assumptions are made in the first week of physics education. A good example is one of the first equations we are taught: the definition of velocity,
\begin{equation}
    v = \frac {\Delta x} {\Delta t}.
\end{equation}
To give this equation precise operational meaning has been an outstanding issue in physics for its entire history. This is because, to understand this equation, one has to have an operational definition of both $x$, $t$, and $\Delta$. Great minds have pondered this question and their insights have led to scientific revolutions. This includes the development of Newtonian mechanics, relativity, and quantum mechanics.\footnote{A lot to digest in the first week!} Recently, the meaning of $x$ and, in particular, $t$, have been the subject of a new debate whose origin is in a theory of quantum gravity. This brings us to our first questionable assumption.

\subsection{Time and space are unified}\label{sec:time and space}

The theory of relativity changed our perception of time. As Minkowski put it in 1908 \cite{Minkowski:seminal_address}, ``space by itself, and time by itself, are doomed to fade away into mere shadows, and only a kind of union of the two will preserve an independent reality.'' Nowhere is this more apparent than in the main equation physicists use to construct the solutions of general relativity (GR):
\begin{equation}\label{eq:EH}
    S_\text{Einstein-Hilbert} =   \int d^4x \left( R  + \mathcal L_\text{matter} \right) \sqrt{-g}    \;.
\end{equation}
Can you spot the $t$? It's hidden in the $4$ of $d^4x$. But there are important structures hidden by this compact notation.



We will start by pointing out an invisible minus sign in equation~\eq{EH}. When calculating spacetime distances, one needs to use
\begin{equation}
    x^2 + y^2 + z^2 - t^2,
\end{equation}
which has a $-$ in front of the $t^2$ instead of Pythagoras' $+$. The minus sign looks innocent but has important consequences for the solutions of equation~\eq{EH}. Importantly, the minus sign implies \emph{causality}, which means that only events in the past can effect what is going on now. This, in turn, implies that generic solutions of GR can only be solved by specifying information at a particular \emph{time} and then seeing how this information propagates into the future. Doing the converse, i.e., specifying information at a particular \emph{place} and seeing how that information propagates to another place, is, in general, not consistent.\footnote{Technically, the difference is in the elliptic versus hyperbolic nature of the evolution equations.} Thus, the minus sign already tells you that you have to use the theory in a way that treats time and space differently.

There are other ways to see how time and space are treated differently in gravity. In Julian Barbour's 2009 essay, \emph{The Nature of Time} \cite{Barbour:nature_of_time}, he points out that Newton's ``absolute'' time is not ``absolute'' at all. Indeed, the Newtonian notion of \emph{duration} -- that is, how much time has ticked by -- can be \emph{inferred} by the total change in the \emph{spatial} separations of particles in the Universe. He derives the equation
\begin{equation}\label{eq:ET}
    \Delta t^2 \propto \sum_i \Delta d^2_i,
\end{equation}
where the $d_i$ are inter--particle separations in units where the masses of the particles are 1. The factor of proportionality is important, but not for our argument. What is important is that changes in time can be inferred by changes in distances so that absolute duration is not an input of the classical theory. This equation can be generalized to gravity where it must be solved at every point in space. The implications for the quantum theory are severe: time completely drops out of the formalism.

Expert readers will recognize this as one of the facets of the \emph{Problem of Time} \cite{Isham:pot_review}. The fact that there is no equivalent \emph{Problem of Space} can be easily traced back to the points just made: time is singled out in gravity as the variable in terms of which the evolution equations are solved. This in turn implies that local duration should be treated as an \emph{inferred} quantity rather than something fundamental. Clearly, time and space are \emph{not} treated on the same footing in the formalism of GR despite the rather misleading form of equation~\eq{EH}. Nevertheless, it is still true that the spacetime framework is incredibly useful and, as far as we know, correct. How can one reconcile this fact with the space--time asymmetry in the formalism itself? We will investigate this in section~(\ref{sec:time from RG}).


\subsection{Scale is physical}\label{sec:scale}

Before even learning the definition of velocity, the novice physicist is typically introduced to an even more primary concept that usually makes up one's first physics lesson: \emph{units}. Despite the rudimentary nature of units, they are probably the most commonly misunderstood concept in all of physics. If you ask ten different physicists for the physical meaning of a unit, you will likely get ten different answers. To avoid confusion, most theoreticians set all dimensionful constants equal to 1. However, one can't predict anything until one has painfully reinserted these dimensionful quantities into the final result.

And yet, no one has \emph{ever} directly observed a dimensionful quantity. This is because all measurements are comparisons. A meter has no intrinsic operational meaning, only the ratio of two lengths does. One can call object A a meter and measure that object B is twice its length. Then, object B has a length of 2 meters but that tells you nothing about the intrinsic length of object A. If a demon doubled the intrinsic size of the Universe, the result of the experiment would be exactly the same. So, where do units come from?

Some units, like the unit of pressure, are the result of emergent physics. We understand how they are related to more ``fundamental'' units like meters and seconds. However, even our most fundamental theories of Nature have dimensionful quantities in them. The standard model of particle physics and classical GR require only a singe unit: \emph{mass}. Scale or, more technically, \emph{conformal invariance} is then broken by the Higgs mass, which is related to all the masses of the particles in the standard model, and the Plank mass, which sets the scale of quantum gravity. As already discussed, there is a naturalness problem associated with writing all other constants of nature as dimensionless quantities.


The presence of dimensionful quantities is an indication that our ``fundamental'' theories are not fundamental at all. Instead, scale independence should be a basic principle of a fundamental theory. As we will see in section~(\ref{sec:time from RG}), there is a formulation of gravity that is \emph{nearly} scale invariant. The ``nearly'' will be addressed by the considerations of the next section.

\subsection{Physical laws are independent of the measurement process}\label{sec:measurement}

There is one assumption that is so fundamental it doesn't even enter the physics curriculum: the general applicability of the scientific method. We know that the scientific method can be applied in the laboratory where external agents (i.e., scientists) carefully control the inputs of some subsystem of the Universe and observe the subsystem's response to these inputs. We don't know, however, whether it is possible to apply these techniques to the Universe as a whole. On the other hand, when it comes to quantum mechanics, we \emph{do} know whether our formalism can be consistently applied to the Universe. The answer is `NO'. The reasons are well understood -- if not disappointingly under appreciated -- and the problem even has a name: \emph{the measurement problem}.

The measurement problem results from the fact that quantum mechanics is a framework more like statistical physics than classical mechanics. In statistical physics, one has \emph{practical} limitations on one's knowledge of a system so one takes an educated guess at the results of a specific experiment by calculating a probability distribution for the outcome using one's current knowledge of the system. In quantum mechanics, one has \emph{fundamental} limitations on one's knowledge of the system -- essentially because of the uncertainty principle -- so one can only make an educated guess at the outcome of a specific experiment by calculating a probability distribution for the outcome using one's current knowledge of the system. However, it would be strange to apply statistical mechanics to the whole Universe\footnote{Believers in the Multiverse could substitute ``Universe'' for ``Multiverse'' in this argument.} because the Universe itself is only given once. It is difficult to imagine an ensemble of Universes for which one can calculate a probability distribution. The same is true in quantum mechanics, but the problem is worse. The framework itself is designed to give you a probability distribution for the outcome of some measurement but how does one even define a measurement when the observer itself is taken to be part of the system? The answer is not found in any interpretation of quantum mechanics, although the problem itself takes a different form in a given interpretation. The truth is that quantum mechanics requires some additional structure, which can be thought of as the observer, in order for it to make sense. In other words, quantum mechanics can never be a theory of the whole Universe.

As a consequence of this, any approach to quantum gravity that uses quantum mechanics unmodified -- including all \emph{major} approaches to quantum gravity -- is not, and \emph{can never be} a theory of the whole Universe. It could still be used for describing quantum gravity effects on isolated subsystems of the Universe, but that is not the ambition of a full fledged quantum gravity theory. Given such a glaring foundational issue at the core of every major approach to quantum gravity, we believe that the attitude that we are nearing the end of physics is unjustified. The ``shut--up and calculate'' era is over. It is time for the quantum gravity community to return to these fundamental issues.

One approach is to change the ambitions of science. This is the safest and most practical option but it would mean that science is inherently a restricted framework. The other possibility is to try to address the measurement problem directly. In the next section, we will give a radical proposal that embraces the role of the observer in our fundamental description of Nature. To understand how this comes about, we need one last ingredient: \emph{renormalization}, or the art of averaging.

\section{A way forward}

\subsection{The art of averaging}

It is somewhat unfortunate that the great discoveries of the first half of the $20^\text{th}$ century have overshadowed those of the second half of the century. One of these, the theory of \emph{renormalization}, is potentially the uncelebrated triumph of $20^\text{th}$ century physics. Renormalization was born as rather ugly set of rules for removing some undesirable features of quantum field theories. From these humble beginnings, it has grown into one of the gems of physics. In its modern form due to Wilson \cite{Wilson:RG_review}, renormalization has become a powerful tool for understanding what happens in a general system when one lacks information about the details of its fine behavior. Renormalization's  reach extends far beyond particle physics and explains, among other things, what happens during phase transitions. But, the theory of renormalization does even more: it helps us understand why physics is possible at all.

Imagine what it would be like if, to calculate everyday physics like the trajectory of Newton's apple, one would have to compute the motions of every quark, gluon, and electron in the apple and use quantum gravity to determine the trajectory. This would be completely impractical. Fortunately, one doesn't have to resort to this. High--school physics is sufficient to determine the motion of what is, fundamentally, an incredibly complicated system. This is possible because one can average, or \emph{coarse grain}, over the detailed behavior of the microscopic components of the apple. Remarkably, the average motion is simple. This fact is the reason why Newtonian mechanics is expressible in terms of simple differential equations and why the standard model is made up of only a couple of interactions. In short, it is why physics is possible at all. The theory of renormalization provides a framework for understanding this.

The main idea behind renormalization is to be able to predict how the laws of physics will change when a coarse graining is performed. This is similar to what happens when one changes the magnification of a telescope. With a large magnification, one might be able to see the moons of Jupiter and some details of the structure of their atmospheres. But, if the magnification, or the \emph{renormalization scale}, is steadily decreased, the resolution is no longer good enough to make out individual moons and the lens averages over these structures. The whole of Jupiter and its moons becomes a single dot. As we vary the renormalization scale, the laws of physics that govern the structures of the system change from the hydrodynamic laws of the atmospheres to Newton's law of gravity.

The theory of renormalization produces precise equations that say how the laws of physics will change, or \emph{flow}, as we change the renormalization scale. In what follows, we will propose that flow under changes of scale may be related to the flow of time.

\subsection{Time from coarse graining}\label{sec:time from RG}

We are now prepared to discuss an idea that puts our three questionable assumptions into a new light by highlighting a way in which they are connected. First, we point out that there is a way to \emph{trade} a spacetime symmetry for conformal symmetry without altering the physical structures of GR. This approach, called \emph{Shape Dynamics} (SD), was initially advocated by Barbour \cite{barbour:bm_review} and was developed in \cite{gryb:shape_dyn,Gomes:linking_paper}. Symmetry trading is allowed because symmetries don't affect the physical content of a theory. In SD, the irrelevance of duration in GR is traded for local scale invariance (we will come to the word ``local'' in a moment). This can be done without altering the physical predictions of the theory but at the cost of having to treat time and space on a different footing. In fact, the local scale invariance is only an invariance of \emph{space}, so that local rods -- not clocks -- can be rescaled arbitrarily. Time, on the other hand, is treated differently. It is a global notion that depends on the total change in the Universe.

In 2 spatial dimensions, we know that this trading is possible because of an accidental mathematical relationship between the structure of conformal symmetry in 2 dimensions and the symmetries of 3 dimensional spacetime \cite{gryb:2_plus_1}.\footnote{Technically, this is the isomorphism between the conformal group in $d$ spatial dimensions and the deSitter group in $d+1$ dimensions.} We are investigating whether this result will remain true in 3 spatial dimensions. If it does, it would mean that the spacetime picture and the conformal picture can coexist because of a mere mathematical accident. 

We now come to a key point: in order for any time evolution to survive in SD, one cannot eliminate all of the scale. The \emph{global} scale of the Universe cannot be traded since, then, no time would flow. Only a \emph{redistribution} of scale from point to point is allowed (this is the significance of the word ``local'') but the overall size of the Universe cannot be traded. In other words, \emph{global scale must remain for change to be possible.} How can we understand this global scale?

Consider a world with no scale and no time. In this world, only 3 dimensional Platonic shapes exist. This kind of world has a technical name, it is a \emph{fixed point} of renormalization -- ``fixed'' because such a world does not flow since the renormalization scale is meaningless. This cannot yet be our world because nothing happens in this world. Now, allow for something to happen and call this ``something'' a \emph{measurement}. One thing we know about measurements is that they can never be perfect. We can only compare the smallest objects of our device to larger objects and coarse grain the rest. Try as we may, we can never fully resolve the Platonic shapes of the fixed point. Thus, coarse graining by real measurements produces flow away from the fixed point. But, what about time? How can a measurement happen if no time has gone by? The scenario that we are suggesting is that the flow under the renormalization scale is exchangeable with the flow of time. Using the trading procedure of SD, the flow of time might be relatable to renormalization away from a theory of pure shape.

In this picture, time and measurement are inseparable. Like a diamond with many faces, scale and time are different reflections of a single entity. This scenario requires a radical reevaluation of our notions of time, scale, and measurement.

To be sure, a lot of thought is still needed to turn this into a coherent picture. A couple of comments are in order. Firstly, some authors \cite{Strominger:holo_cosmo,Skenderis:holo_uni} have investigated a similar scenario, called \emph{holographic cosmology} using something called \emph{gauge/gravity duality}. However, our approach suggests that one may not have to \emph{assume} gauge/gravity duality for this scenario but, instead, can make use of symmetry trading in SD. Furthermore, our motivation and our method of implementation is more concrete. Secondly, why should we expect that there is enough structure in a coarse graining of pure shapes to recover the rich structure of spacetime? A simple answer is the subject of the next section.\footnote{FM and M. Lostaglio are exploring a related approach \cite{Matteo:thesis}.}

\section{The size that matters}

In this section, we perform a simple calculation suggesting that the coarse graining of shapes described in the last section could lead to gravity. This section is more technical than the others but this is necessary to set up our final result. Brave souls can find the details of the calculations in the Technical Appendix~(\ref{TechnicalAppendix}).

\begin{figure}
\begin{center}\includegraphics[width=0.75\textwidth]{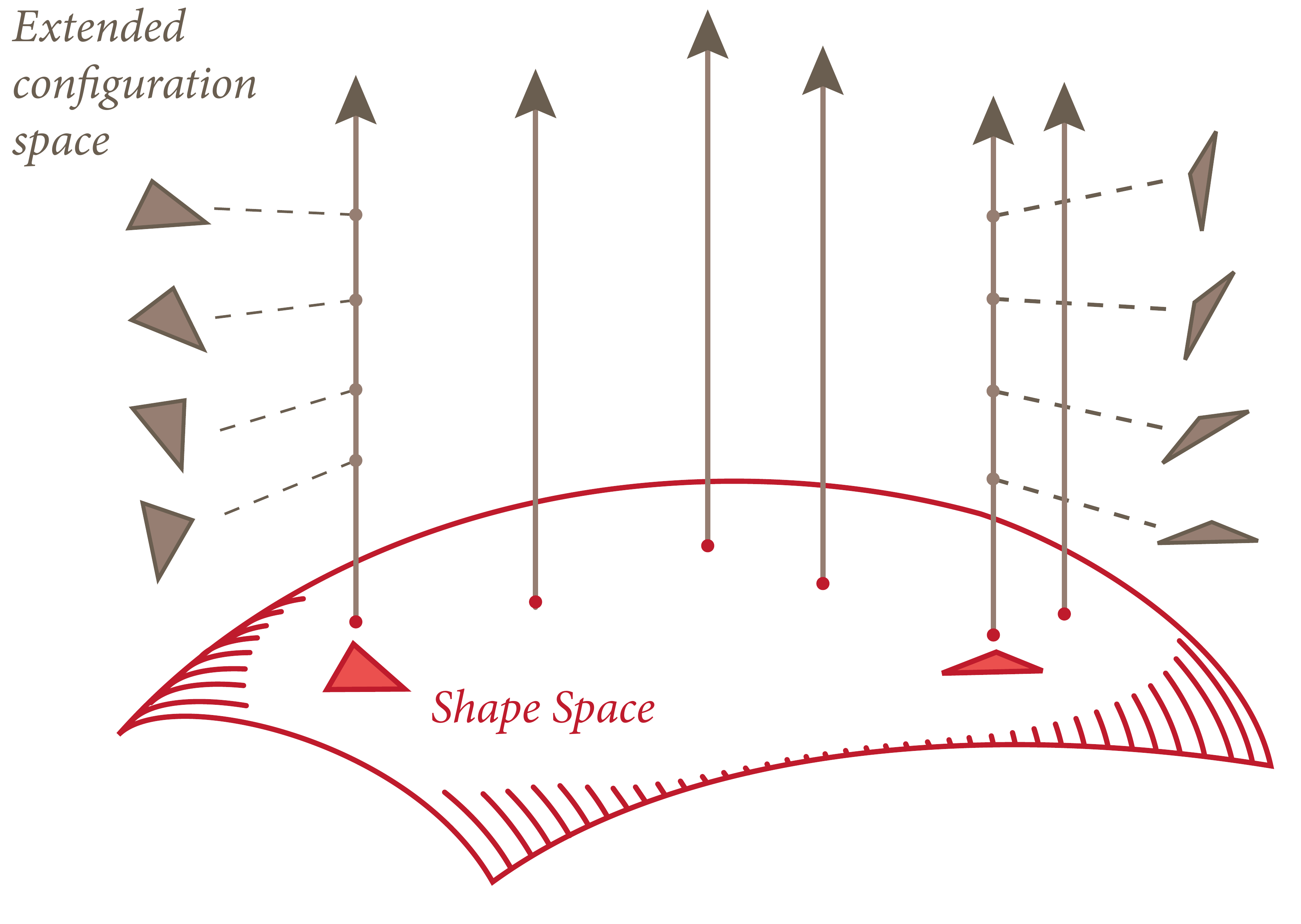}\end{center}
\caption{Each point in Shape Space is a different shape (represented by triangles). These correspond to an equivalence class (represented by arrows) of points of the Extended Configuration Space describing the same shape with a different position, orientation, and size.} \label{pfb}
\end{figure}

We will consider a simple ``toy model'' that, remarkably, recovers a key feature of gravity. Our model will be a set of $N$ free Newtonian point particles. To describe the calculation we will need to talk about two spaces: \emph{Shape Space} and \emph{Extended Configuration Space} (ECS). Shape Space is the space of all the shapes of the system. If $N=3$, this is the space of all triangles. ECS is the space of all Cartesian coordinates of the particles. That is, the space of all ways you can put a shape into a Cartesian coordinate system. The ECS is larger than Shape Space because it has information about the position, orientation, and size of the shapes. Although this information is unphysical, it is convenient to work with it anyway because the math is simpler. This is called a \emph{gauge theory}. We can work with gauge theories provided we remove, or \emph{quotient}, out the unphysical information. To understand how this is done, examine Figure~(\ref{pfb}) which shows schematically the relation between the ECS and Shape Space. Each point on Shape Space is a different shape of the system, like a triangle.
All the points along the arrows represent the same shape with a different position, orientation, or size. By picking a representative point along each arrow, we get a 1--to--1 correspondence between ECS and Shape Space. This is called \emph{picking a gauge}. Mathematically, this is done by imposing constraints on the ECS. In our case, we need to specify a constraint that will select a triangle with a certain center of mass, orientation, and size. For technical reasons, we will assume that all particles are confined to a line so that we don't have to worry about orientation. To specify the size of the system, we can take the ``length'' of the system, $R$, on ECS. This is the \emph{moment of inertia}. By fixing the center of mass and moment of inertia in ECS, we can work indirectly with Shape Space. The main advantage of doing this is that there is a natural notion of distance in ECS. This can be used to define the distance between two shapes, which is a key input of our calculations.

To describe the calculation, we need to specify a notion of \emph{entropy} in Shape Space. Entropy can be thought of as the amount of information needed to specify a particular macroscopic state of the system. To make this precise, we can use the notion of distance on ECS to calculate a ``volume'' on Shape Space. This volume roughly corresponds to the number of shapes that satisfy a particular property describing the state. The more shapes that have this property, the more information is needed to specify the state. The entropy of that state is then related to its volume, $\Omega_m$, divided by the total volume of Shape Space, $\Omega_\text{tot}$. Explicitly,
\begin{equation}
    S = -k_\text{B} \log \frac{\Omega_m}{\Omega_\text{tot}},
\end{equation}
where $k_\text{B}$ is Boltzmann's constant.

\begin{figure}
\begin{center}\includegraphics[width=0.75\textwidth]{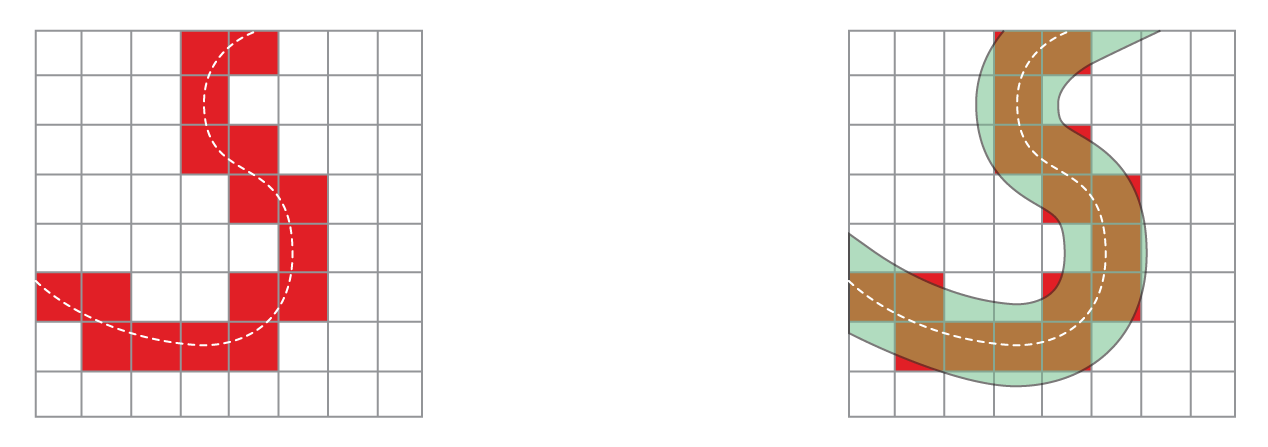}\end{center}
\caption{\emph{Left:} Approximation of a line using a grid.
\emph{Right:} Further approximation of the line as a strip of thickness equal to the grid spacing.} \label{LatticeApprox}
\end{figure}

We will be interested in states described by a subsystem of $n<N$ particles that have a certain center of mass ${\bm x}_0$ and moment of inertia, $r$. To make sense of the volume, we need a familiar concept: coarse graining. We can approximate the volume of the state by chopping up the ECS into a grid of size $\ell$. Physically, the coarse graining means that we have a measuring device with a finite resolution given by $\ell$. Consider a state that is represented by some surface in ECS. This is illustrated in Figure~(\ref{LatticeApprox}) by a line.
The volume of the state is well approximated by counting the number of dark squares intersected by the line. In the Technical Appendix~(\ref{TechnicalAppendix}), we calculate this volume explicitly. The result is
\begin{equation}
    \Omega_\text{m} \propto \ell^2 \; r^{n-2} \; \left( R^2 - r^2 - \left(1 + \frac m {M-m} \right)\frac m M \; x_0^2\right)^{\frac{N-n-2}{2}} \;,
\end{equation}
where $M$ and $R$ are the total mass and moment of inertia of the whole system and $m$ is the mass of the subsystem. We can then compare this volume to the total volume of Shape Space, which goes like the volume of an $N-1$ dimensional sphere (the $-1$ is because of the center of mass gauge fixing). Thus,
\begin{equation}
    \Omega_\text{tot} \propto R^{N-1}.
\end{equation}
The resulting entropy is
\begin{equation}\label{eq:entropy}
    S = \frac 1 2 \, k_B \, \frac N n \, \left( \frac r R \right)^2 -  \, k_B \, \log  \frac r R + \dots.
\end{equation}
Remarkably, the first term is exactly the entropy of a black hole calculated by Bekenstein and Hawking \cite{BCH,Bekenstein}. More remarkably, the second term is exactly the first correction to the Bekenstein--Hawking result calculated in field theory \cite{BlackHole1,BlackHole2}. Erik Verlinde \cite{Verlinde:entropic_gravity} discovered a way to interpret Newtonian gravity as an \emph{entropic} force for systems whose entropy behaves in this way. It would appear that this simple model of a coarse graining of pure shapes has the right structure to reproduce Newtonian gravity.

\section{Conclusions}

We have questioned the basic assumptions that: i) time and space should be treated on the same footing, ii) scale should enter our fundamental theories of Nature, and iii) the evolution of the Universe is independent of the measurement process. This has led us to a radical proposal: that time and scale emerge from a coarse graining of a theory of pure shape. The possibility that gravity could come out of this formalism was suggested by a simple toy model. The results of this model are non--trivial. The key result was that the entropy \eq{entropy} scales like $r^2$, which, dimensionally, is an area. In three dimensions, this is the signature of \emph{holography}. Thus, in this simple model, Shape Space is holographic. If this is a generic feature of Shape Space, it would be an important observation for quantum gravity.

Moreover, the toy model may shed light on the nature of the Plank length. In this model, the Plank length is the emergent length arising in ECS given by
\begin{equation}
    L_\text{Planck}^2 = G \, \hbar \propto \frac {R^2} N \;.
\end{equation}
This dimensionful quantity, however, is not observable in this model. What is physical, instead, it the dimensionless ratio $r/R$. This illustrates how a dimensionful quantity can emerge from a scale independent framework. Size doesn't matter -- but a ratio of sizes does. The proof could be gravity.

\clearpage
\appendix

\section{Technical Appendix}
\label{TechnicalAppendix}

The extended configuration space is $\mathbbm R^N$: the space coordinates, $r_i$, ($i= 1 ,\dots , N$)
of $N$ particles in 1 dimension. To represent the reduced configuration space, or Shape Space, we can use a
gauge fixing surface. To fix the translations, we can fix the center of mass to be at the origin of the coordinate system:
\begin{equation}
\sum_{i=1}^{N} \, m_i \; r_i = 0 \;. \qquad \text{\it(center of mass at the origin)} \label{TotCenterOfMass}
\end{equation}
The equation above gives three constraints selecting three orthogonal planes through the origin whose orientation is determined by the masses $m_i$.
A natural gauge--fixing for the generators of dilatations is to set the moment of inertia with respect to the center of mass to a constant\footnote{We are using here the notion
of moment of inertia with respect to a point, which we rescaled by
the total mass $M = \sum_i m_i$ to give it the dimensions of a squared length.} 
(the weak equation holds when the gauge--fixing (\ref{TotCenterOfMass}) is applied):
\begin{equation} \label{DilatationGaugeFixing}
\sum_{i<j} \, \frac{m_i m_j}{M^2} \,| r_i - r_j|^2 \approx 
\sum_{i=1}^{N} \, \frac{m_i}{M} \; | r_i |^2= R^2\;. \qquad \text{\it(fixed moment of inertia)}
\end{equation}
The last relation defines a sphere in $\mathbbm R^N$ centered at the origin. Thus, Shape Space is the intersection of the $N-1$-dimensional sphere (\ref{DilatationGaugeFixing}) with the three orthogonal planes (\ref{TotCenterOfMass}).

The flat Euclidean metric, $ ds^2 = m_i \; \delta_{ij} \; \delta_{ab}  \; d r_i^a \; d  r_j^b $, is the natural metric on the extended configuration space $Q$. This metric induces the non--flat metric
\begin{equation}
ds^2_\text{induced} = \left. m_i \; \delta_{ij} \; \delta_{ab}  \; d r_i^a \; d  r_j^b  \right|_{Q_S} \;.
\label{InducedMetric}
\end{equation}
on Shape Space.

\subsection{Description of a macrostate in Shape Space}

Consider an $N$--particle toy Universe with an $n$--particle subsystem, $n<N$.
The particles in the subsystem have coordinates $x_i =  r_i$,  ($i = 1,\dots, n$),
while the coordinates of all the other particles will be called $y_i = r_{n+i}$ ,
($i= 1,\dots,N-n$). It is useful to define the coordinates of the center of mass of the subsystem
and of the rest of the Universe:\footnote{Notice that the two sets of coordinates must satisfy
the relation $ m \; x_0 + (M-m) y_0 = 0$ in order to keep the total center of mass
at the origin.}
\begin{equation} \label{CentersOfMass}
x_0 = \sum_{i=1}^n \frac{m_i}{m} \, x_i  \;, \qquad  y_0 = \sum_{i=1}^{N-n} \frac{m_{n+i}}{M-m} \, y_i  \;, \qquad  ~~~  m = \sum_{i=1}^n m_i\;,
\end{equation}
and the center--of--mass moment of inertia of the two subsystems
\begin{equation}\label{MomentsOfInertia}
r = \sum_{i=1}^n \frac{m_i}{M} \, | x_i - x_0 |^2   \;, \qquad r' = \sum_{i=1}^{N-n} \frac{m_{n+i}}{M} \,| y_i -  y_0 |^2  \;.
\end{equation}
The relation between the moments of inertia of the total system and those of the two
subsystems is 
\begin{equation}
R^2 = r^2 + (r')^2 + \left( 1 + \frac m {M-m} \right) \frac m M \; x_0^2 \;. \label{TotalMomentOfInertia}
\end{equation}

We define a macrostate as a state in which the moment of inertia of the subsystem, $r$,
and its center of mass, ${\bm x}_0$, are constant.
To calculate the Shape Space volume of such a macrostate, we must integrate
over all Shape Space coordinates ${\bm x}_i$ and ${\bm y}_i$ that respect the conditions (\ref{CentersOfMass}),
(\ref{MomentsOfInertia}), and (\ref{TotalMomentOfInertia})
using the measure provided by the induced metric (\ref{InducedMetric}). Let's make the following
change of variables:
\begin{equation}
 {\tilde  x}_i = \sqrt{m_i}  \left( x_i -  x_0 \right)  \;,
\qquad
{\tilde y}_i =  \sqrt{m_{n+i}}  \left( y_i  -  y_0 \right)  \;.
\end{equation}
Our equations become
\begin{equation}
\begin{array}{c}
\frac 1 m \sum_{i=1}^n \sqrt{m_i} \; {\tilde x}_i  = 0  \;, ~~~  \frac 1 {M-m} \sum_{i=1}^n \sqrt{m_{n+i}} \; {\tilde y}_i  = 0
\;,\\\\
r = \frac{1}{M} \sum_{i=1}^n {\tilde x}_i^2   \;, ~~~ r' = \frac 1 {M} \sum_{i=1}^{N-n} {\tilde y}_i^2  \;, ~~~
R^2 = r^2 + (r')^2 + \left( 1 + \frac m {M-m} \right) \frac m M \; x_0^2 \;.
\end{array}
\end{equation}
In the new coordinates, the metric is the identity matrix (it loses the $m_i$ factors on the diagonal). The integral is over the direct product of an $(n-2)$--dimensional sphere of radius $Mr$ and an $(N-n-2)$--dimensional sphere of radius $Mr' = M \sqrt{R^2 - r^2 - \left( 1 + \frac m {M-m} \right) \frac m M \; x_0^2}$ whose volume (calculated with a
coarse--graining of size $\ell$) is:
\begin{equation}
\Omega_\text{m} = \ell^2 \frac{ 4 \; \pi^{(N-n-1)/2} \pi^{(n-1)/2} }{ \Gamma((N-n-1)/2) \Gamma((n-1)/2)} M^{N-4}  r^{n-2} \left(R^2 - r^2 - \left( 1 + \frac m {M-m} \right) \frac m M \; x_0^2 \right)^{\frac {N-n-2}{2}}\;.
\end{equation}
The total volume of Shape Space is that of an $(N-1)$--dimensional sphere of radius $M R$
\begin{equation}
\Omega_\text{tot} =\frac{ 2 \pi^{N/2} }{ \Gamma(N/2)} \; M^{N-1} \, R^{N-1} \;.
\end{equation}
Thus, the Shape Space volume per particle, in the limit $1 \ll n \ll N$, $r \ll r$, $m \ll M$ reduces to
\begin{equation}
\omega \propto \left( \frac \ell r \right)^{2/n} \; \frac r R  \left(1 - \left( \frac r R \right)^2 - \left( 1 + \frac m {M-m} \right) \frac m M \; \left( \frac{x_0} R \right)^2 \right)^{\frac {N}{2n}} \;,
\end{equation}
and its logarithm has the expansion (remember that $x_0 < R$) 
\begin{equation}
S =  \frac 1 2  k_B \frac {N}{n}  \left( \frac r R \right)^2  - k_B \log \frac r R  - \frac 2 n k_B \; \log  \frac \ell r  + \dots \;.
\end{equation}
Notice that the numerical factors change in the 3 dimensions. In that case, they are
\begin{equation}
S = \frac 3 2 \, k_B \, \frac N n \, \left( \frac r R \right)^2 -  3 \, k_B \, \log  \frac r R - \frac 4 n \, k_B \log \frac \ell r \dots \;.
\end{equation}


\clearpage
\bibliographystyle{utphys}
\bibliography{mach,BibEssay}

\providecommand{\href}[2]{#2}\begingroup\raggedright\begin{thebibliography}{10}

\bibitem{Kelvin:dark_clouds}
L.~Kelvin, ``Nineteenth Century Clouds over the Dynamical Theory of Heat and
  Light,'' {\em Philosophical Magazine, Sixth Series} {\bfseries 2} (1901)
  1--40. From a 1900, April 27, Royal Institution lecture.

\bibitem{Minkowski:seminal_address}
H.~Minkowski, {\em The Principle of Relativity: A Collection of Original
  Memoirs on the Special and General Theory of Relativity}, ch.~Space and Time,
  pp.~75--91.
\newblock New York: Dover, 1952.

\bibitem{Barbour:nature_of_time}
J.~Barbour, ``{The Nature of Time},''
  \href{http://arxiv.org/abs/0903.3489}{{\ttfamily arXiv:0903.3489 [gr-qc]}}.

\bibitem{Isham:pot_review}
C.~J. Isham, ``{Canonical quantum gravity and the problem of time},''
  \href{http://arxiv.org/abs/gr-qc/9210011}{{\ttfamily arXiv:gr-qc/9210011}}.

\bibitem{Wilson:RG_review}
K.~Wilson and J.~B. Kogut, ``{The Renormalization group and the epsilon
  expansion},'' \href{http://dx.doi.org/10.1016/0370-1573(74)90023-4}{{\em
  Phys.Rept.} {\bfseries 12} (1974) 75--200}.

\bibitem{barbour:bm_review}
J.~Barbour, ``Dynamics of pure shape, relativity and the problem of time,'' in
  {\em Decoherence and Entropy in Complex Systems}, Springer Lecture Notes in
  Physics.
\newblock 2003.
\newblock Proceedings of the Conference DICE, Piombino 2002, ed. H.-T Elze.

\bibitem{gryb:shape_dyn}
H.~Gomes, S.~Gryb, and T.~Koslowski, ``{Einstein gravity as a 3D conformally
  invariant theory},''
  \href{http://dx.doi.org/10.1088/0264-9381/28/4/045005}{{\em Class. Quant.
  Grav.} {\bfseries 28} (2011) 045005},
  \href{http://arxiv.org/abs/1010.2481}{{\ttfamily arXiv:1010.2481 [gr-qc]}}.

\bibitem{Gomes:linking_paper}
H.~Gomes and T.~Koslowski, ``{The Link between General Relativity and Shape
  Dynamics},'' {\em Class.Quant.Grav.} {\bfseries 29} (2012) 075009,
  \href{http://arxiv.org/abs/1101.5974}{{\ttfamily arXiv:1101.5974 [gr-qc]}}.

\bibitem{gryb:2_plus_1}
S.~Gryb and F.~Mercati, ``{2+1 gravity on the conformal sphere},''
  \href{http://arxiv.org/abs/1209.4858}{{\ttfamily arXiv:1209.4858 [gr-qc]}}.

\bibitem{Strominger:holo_cosmo}
A.~Strominger, ``{Inflation and the dS / CFT correspondence},'' {\em JHEP}
  {\bfseries 0111} (2001) 049,
  \href{http://arxiv.org/abs/hep-th/0110087}{{\ttfamily arXiv:hep-th/0110087
  [hep-th]}}.

\bibitem{Skenderis:holo_uni}
P.~McFadden and K.~Skenderis, ``{The Holographic Universe},''
  \href{http://dx.doi.org/10.1088/1742-6596/222/1/012007}{{\em
  J.Phys.Conf.Ser.} {\bfseries 222} (2010) 012007},
  \href{http://arxiv.org/abs/1001.2007}{{\ttfamily arXiv:1001.2007 [hep-th]}}.

\bibitem{Matteo:thesis}
F.~Mercati and M.~Lostaglio, ``Scale Anomaly as the origin of Time,''. In
  preparation.

\bibitem{BCH}
J.~M. Bardeen, B.~Carter, and S.~W. Hawking, ``The four laws of black hole
  mechanics,'' \href{http://dx.doi.org/10.1007/BF01645742}{{\em Commun. Math.
  Phys.} {\bfseries 31} (1973) 161--170}.

\bibitem{Bekenstein}
J.~D. Bekenstein, ``Black holes and entropy,''
  \href{http://dx.doi.org/10.1103/PhysRevD.7.2333}{{\em Phys. Rev.} {\bfseries
  7} (1973) 2333--2346}.

\bibitem{BlackHole1}
K.~S. Gupta and S.~Sen, ``{Further evidence for the conformal structure of a
  Schwarzschild black hole in an algebraic approach},''
  \href{http://dx.doi.org/10.1016/S0370-2693(01)01501-5}{{\em Phys.Lett.}
  {\bfseries B526} (2002) 121--126},
  \href{http://arxiv.org/abs/hep-th/0112041}{{\ttfamily arXiv:hep-th/0112041}}.

\bibitem{BlackHole2}
D.~Birmingham and S.~Sen, ``{An Exact black hole entropy bound},''
  \href{http://dx.doi.org/10.1103/PhysRevD.63.047501}{{\em Phys.Rev.}
  {\bfseries D63} (2001) 047501},
  \href{http://arxiv.org/abs/hep-th/0008051}{{\ttfamily arXiv:hep-th/0008051}}.

\bibitem{Verlinde:entropic_gravity}
E.~P. Verlinde, ``{On the Origin of Gravity and the Laws of Newton},''
  \href{http://dx.doi.org/10.1007/JHEP04(2011)029}{{\em JHEP} {\bfseries 04}
  (2011) 029}, \href{http://arxiv.org/abs/1001.0785}{{\ttfamily arXiv:1001.0785
  [hep-th]}}.

\end{thebibliography}\endgroup

\end{document}